# Simple model of skeletal matter
# composed of magnetized electrically conducting thin rods

A.B. Kukushkin, K.V. Cherepanov

*NFI RRC "Kurchatov Institute", Moscow, 123182, Russia*

A simple electrodynamic model for describing the behavior of a skeletal matter composed of magnetized, electrically conducting thin rods (1D magnetic dipoles) is proposed. It is aimed at modeling the self-assembling of a skeletal matter from carbon nanotubes (or similar nanodust), as suggested in [1] for interpreting the experimental data on the long-lived filamentary structures in the high-current electric discharges. Here the capability of the model is illustrated with the example of how a straight tubular skeleton, which is composed of ~300 dipoles and carry circular electric current in its wall, may be wrapped up by a distant pulsed electric current to make a toroid-like structure.

## 1. Introduction

The self-similar skeletal structures [1,2], composed of tubular blocks (sometimes with the cartwheel in the butt-end of the tubule) which repeat themselves successively at various length scales to give, correspondingly, a fractal of particular topology of constituent blocks, were called the Universal Skeletal Structures (USS) [3]. The phenomenon of USS was suggested in [1] for interpreting the experimental data on the long-lived filamentary structures in the high-current electric discharges. The USS phenomenon was predicted and traced in the very wide range of length scales, $10^{-5}$ cm - $10^{23}$ cm, in the data from various laboratory experiments and observations of severe weather phenomena and astrophysical objects [2] (for the current status of the USS project and its popular description see, respectively, [4,5] and [3]).

The smallest block of USS was suggested [1] to be the widely known object, namely carbon nanotube, or similar nanostructures with participation of other chemical elements. The prediction [1] was based on appealing to exceptional electrodynamic properties of their hypothetical building blocks -- first of all, the ability of these blocks to facilitate the electric breakdown in laboratory discharges and to assemble the micro- and macroskeletons. The self-assembling of skeletons was suggested to be based dominantly on *magnetic* phenomena. The indications on plausibility of the anomalous magnetism and, in particular, on the ability of CNTs, and/or their assemblies, to trap and almost dissipationlessly hold magnetic flux, with the specific magnetization high enough to stick the CNTs together, come from observations of superconductor-like diamagnetism in the assemblies of CNTs at high enough temperatures. Such evidences are obtained for the self-assemblies of CNTs (which contain, in particular, the ring-shaped structures of few tens of microns in diameter) inside *non-processed* fragments of cathode deposits, at room temperatures, [6] and for the artificial assemblies, at 400 K [7]. The evidences and arguments for the room-temperature superconductivity in individual CNT, and in artificial and natural assemblies of CNTs, are summarized in [8]. The recent survey of experimental evidences for, and theoretical models of, the unexpected magnetism of carbon foams and heterostructured nanotubes is given in [9].

Despite the above experimental evidences and theoretical models need much stronger tests and confirmations, they justify explicit demonstration of the capability of magnetized nanotubular blocks to self-assemble a tubule of higher generation [1(B,C)] and sustain the integrity of the assembled skeleton. Similarly to development of, e.g., the plasma theory, now it is worth to start with analysing the stability of skeletal matter within the frame of as simple model as possible. This implies an analysis of the capability of nanotubes to sustain the integrity of the hypothetically formed tubular skeleton, which is composed of magnetized nanotubes (i.e. analyse the viability of the tubule of the 2-nd generation).

In the present paper, we (i) formulate a simple model for describing the behavior of a skeletal matter composed of magnetized, electrically conducting thin rods which behave as the 1D magnetic dipoles, and (ii) illustrate the capability of the model on the example of how a straight tubular skeleton, which is composed of ~300 dipoles and carry circular electric current in its wall, may be wrapped up by a distant pulsed electric current to make a toroid-like structure.

## 2. Simple model of skeletal matter, composed of 1D magnetic dipoles

We treat the problem in as simple picture as possible. Thus, we assume the elementary block of the skeletons to possess the following electrodynamic properties:
- the 1D static magnetic dipole (such a dipole may be represented as a couple of magnetic monopoles located on the tips of the rigid-body dipole; this approximation seems to be good for the tubules and/or rod with the large length-to-diameter ratio),
- static positive electric charge, which is located in the center of the rod and is screened by the ambient electrons at some Debye radius (electric charging is due to inevitable field emission, at least thermal one, by the nanotubes),
- static electrical conductivity, which is high enough to enable the tubular skeleton to trap, without dissipation, the magnetic flux inside the tubule (i.e. sustain circular electric currents in the tubule's wall).

The above characteristics enable us to describe the following interactions of elementary blocks:
- mutual magnetic attraction and repulsion of the dipoles (i.e. interaction of circular electric current in the wall of one elementary block with similar current in another elementary block),
- action of external magnetic field on the magnetic dipole (i.e. interaction of circular electric currents in the walls of the elementary block with the external electric current producing the magnetic field),
- screened electric repulsion of elementary blocks,
- action of magnetic field, produced by the longitudinal electric current in all the magnetic dipoles, on the given dipole (i.e. interaction of circular electric current in the walls of the elementary block with the longitudinal electric current in the walls of other blocks),
- interaction of longitudinal electric current in the walls of the blocks.

To simplify the description of dynamics of solid bodies we consider each dipole to be a couple of the point objects (coordinates $r_i$, masses $m_i$, i=1,2) which are linked together with a rigid-body massless bond and subjected to the action of the external forces applied to these objects, $F_1$ and $F_2$, and to the center of mass of the system (i.e. to the massless interconnecting bond), $F_{cm}$. The exact system of equations for such a system is described by the conventional equations for the motion of a solid body specified for the above particular case. The equations for the momentum and angular momentum, respectively, of the solid body are as follows:

$$\frac{\partial^2}{\partial t^2}(m_1 \vec{r}_1 + m_2 \vec{r}_2) = \vec{F}_1 + \vec{F}_2 + \vec{F}_{cm}. \tag{1}$$

$$m_1 \left[ \vec{r}_1, \frac{\partial^2}{\partial t^2} \vec{r}_1 \right] + m_2 \left[ \vec{r}_2, \frac{\partial^2}{\partial t^2} \vec{r}_2 \right] = \left[ \vec{r}_1, \vec{F}_1 \right] + \left[ \vec{r}_2, \vec{F}_2 \right] + \left[ \vec{r}_{cm}, \vec{F}_{cm} \right] \tag{2}$$

where the square brackets denote the vector multiplication, and the radius vector of the center of mass of the system is equal to

$$\vec{r}_{cm} \equiv (m_1 \vec{r}_1 + m_2 \vec{r}_2)/(m_1 + m_2). \tag{3}$$

One may easily check that the solution to the system of Eqs. (1) and (2) may be found as a solution to the following system of equations:

$$m_1 \frac{\partial^2}{\partial t^2} \vec{r}_1 = -A \vec{r}_{12} + \vec{F}_1 + \frac{m_1}{m_1 + m_2} \vec{F}_{cm}. \tag{4}$$

$$m_2 \frac{\partial^2}{\partial t^2} \vec{r}_2 = A \vec{r}_{12} + \vec{F}_2 + \frac{m_2}{m_1 + m_2} \vec{F}_{cm}. \tag{5}$$

$$\vec{r}_{12} \equiv (\vec{r}_1 - \vec{r}_2) \tag{6}$$

The first term in the right-hand side of Eqs. (4) and (5) describes the action of the rigid-body bond between the point objects 1 and 2. One can find the value of $A$ from the condition of rigidity of a solid body,

$$(\vec{v}_1 - \vec{v}_2, \vec{r}_1 - \vec{r}_2) \equiv (\vec{v}_{12}, \vec{r}_{12}) = 0, \tag{7}$$

where $\vec{v}_1$ and $\vec{v}_2$ are the velocities of the point objects. This gives

$$A = \frac{\mu_{12}}{r_{12}^2} \left\{ v_{12}^2 + \left( \frac{\vec{F}_1}{m_1} - \frac{\vec{F}_2}{m_2}, \vec{r}_{12} \right) \right\}, \tag{8}$$

where $\mu_{12}$ is the reduced mass of the system of two point masses.

Major dimensionless variables of the outlined above problem are as follows. The space coordinates, time and velocity are taken in the units of dipole's length $L$, $t_0$ and $v_0$, respectively:

$$r_0 = L, \quad t_0 = \frac{\sqrt{mL^3}}{Z_M e}, \quad v_0 = \frac{Z_M e}{\sqrt{mL}}, \quad Z_M = \frac{\Phi_0}{4\pi e}, \qquad (9)$$

where $m=m_1=m_2$, $Z_M$ is the modulus of magnetic charge of the monopole taken in the units of electron charge $e$, $\Phi_0$ is magnetic flux trapped in the dipole. Electric charge Z will be taken in the units of magnetic charge.

All the forces are expressed in the units of magnetic interaction attraction at the distance $L$. The pair interaction of longitudinal electric currents of the value $J_o$ through the dipole is taken in the units of $F_{0JJ}$, and the interaction of the dipoles with external current $J_{ext}$ - in the units of $F_{Jext}$:

$$F_{0JJ} = \left(\frac{J_0 L}{cZ_M e}\right)^2, \qquad F_{Jext} = \frac{J_{ext} L}{cZ_M e} \qquad (10)$$

The electrodynamic forces are assumed to largely exceed the gravity of the dipoles.

To describe sticking of the dipoles we allow the magnetic monopoles to move freely in an isotropic potential well which is formed by (a) magnetic attraction of monopoles of the opposite sign and (b) their repulsion due to elasticity of the tips of the tubules/rods of finite diameter. The form of the potential and the respective force are shown in Figure 1. This potential provides smooth transition from the Coulomb potential for r>r* to repulsion potential at small radii. Also, in the region r<r* we introduced the following friction force:

$$\vec{F}_{brake} = k_{br} \vec{v}_{12} v_{12}, \qquad (11)$$

where $v_{12}$ is the relative velocity, and the coefficient $k_{br}$ is taken in the units $m/L$.

The above strong simplification of the original picture of the motion of solid rods is acceptable if the spatial density of the rods is rather small and, respectively, the sticking and collisions of the rods are governed mostly by the interaction of strong magnetic monopoles on the tips of these rods.

## 3. Dynamics of tubular skeleton, composed of 1D magnetic dipoles

Here we illustrate the capability of the model, outlined in the previous Section, to describe the integrity of skeleton under the action of external forces. First, we construct the ideal tubular skeleton according to the rules suggested in [1(B,C)]: namely, the skeleton is composed of hexagons assembled from the dipoles. The structure of the tubular straight skeleton, whose wall is assembled from hexagons and whose cross section has also a hexagonal structure, is shown in Figs. 2 and 3, for the total number of the dipoles $N_{dip}$ = 294. The corresponding magnetic threading of such a network will be ideal if the skeleton is composed of the dipoles of magnetic charges $Z_M$ which differ by the factor of 2. In general case, it is possible to compose a skeleton from arbitrary polygons provided the magnetic charges on the tips of the blocks support the respective magnetic threading.

The dynamics of the skeleton in Figs. 2,3 is tested against the perturbation introduced by the distant external electric current for the following conditions,:
- magnetic charges $Z_M=2$, for red thick blocks, and $Z_M=2$, for all the others,
- electric charges $Z=1$ for all the blocks,
- screening (Debye) radius $r_D=1$,
- brake coefficient  $k_{br}=100$,
- current-current force coefficient $F_{0JJ}=2$,
- current-external-current force coefficient $F_{Jext}=50$,
- external electric current flows along X-direction, the line of current is located in the point {Y=-15, Z=15} and acts from time t=0 to t=1.

The results of numerical modeling are shown in Figures 4-7 for various time moments. The future dynamics of the skeleton -- collision of the tips of the skeleton, which follows the "closure" of the loop, as is seen in Figure 7 -- may not be described by the model of Sec. 2 because we neglected mechanical collision of the rods along entire length of the blocks.

The results of Figs. 4-7 may be interpreted as an illustration of the possibility of skeletons -- if formed in the high-current electric discharges or similar conditions - to form the torodal-like and cartwheel-like structures (cf. laser-induced production of large carbon-based toroids reported in [10], see the Q-shaped toroids in Fig. 3 of this paper).

**Acknowledgments**

One of the authors (A.B.K.) highly appreciates his long-term collaboration with V.A. Rantsev-Kartinov in their research of skeletal structures [1-5].
The present work is supported by the Russian Foundation for Basic Research (project No. 05-08-65507).

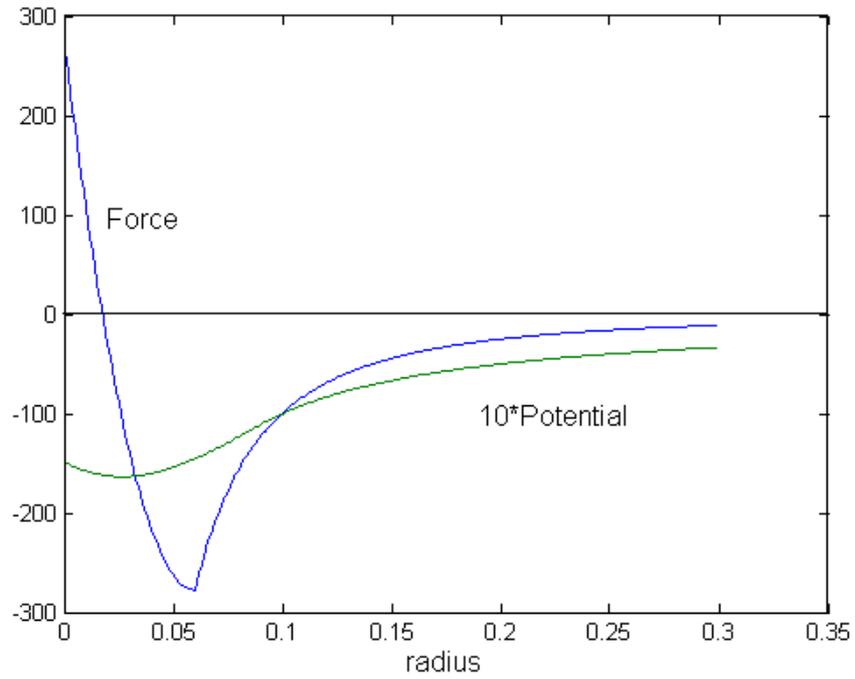

Figure 1. Radial dependence of the effective model potential (taken in the units of $(Z_M e)^2/L$, and multiplied by 10) and the respective force (in the units $(Z_M e/L)^2$) for the interaction of two attracting magnetic monopoles. Here, transition radius is $r^* = 0.06$.

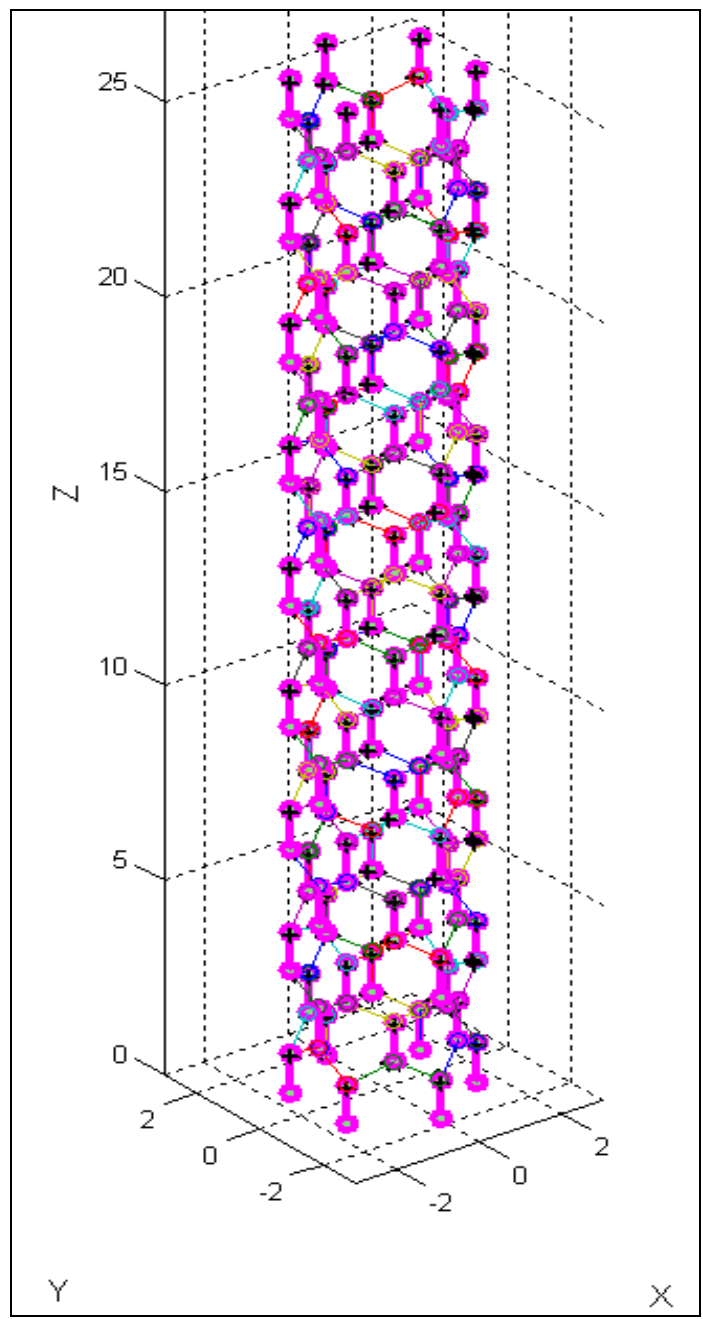

Figure 2. Tubular skeletal structure composed of 294 magnetic dipoles. Magnetic charge of the dipoles shown as red thick rods is twice of that for thin rods.

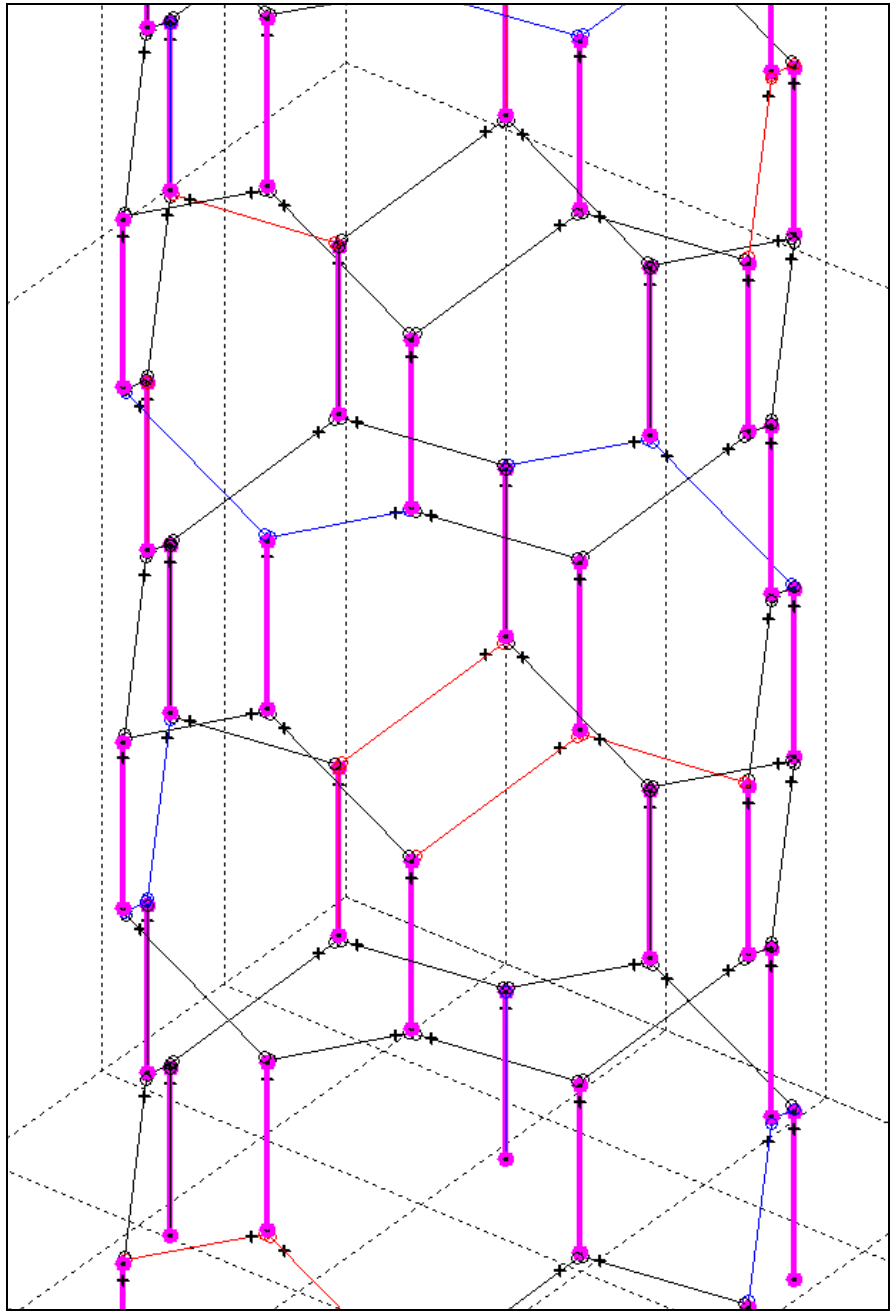

Figure 3. Magnified image of the part of tubular skeleton in Fig. 2. The crosses on the dipoles indicate north pole part of the dipole.

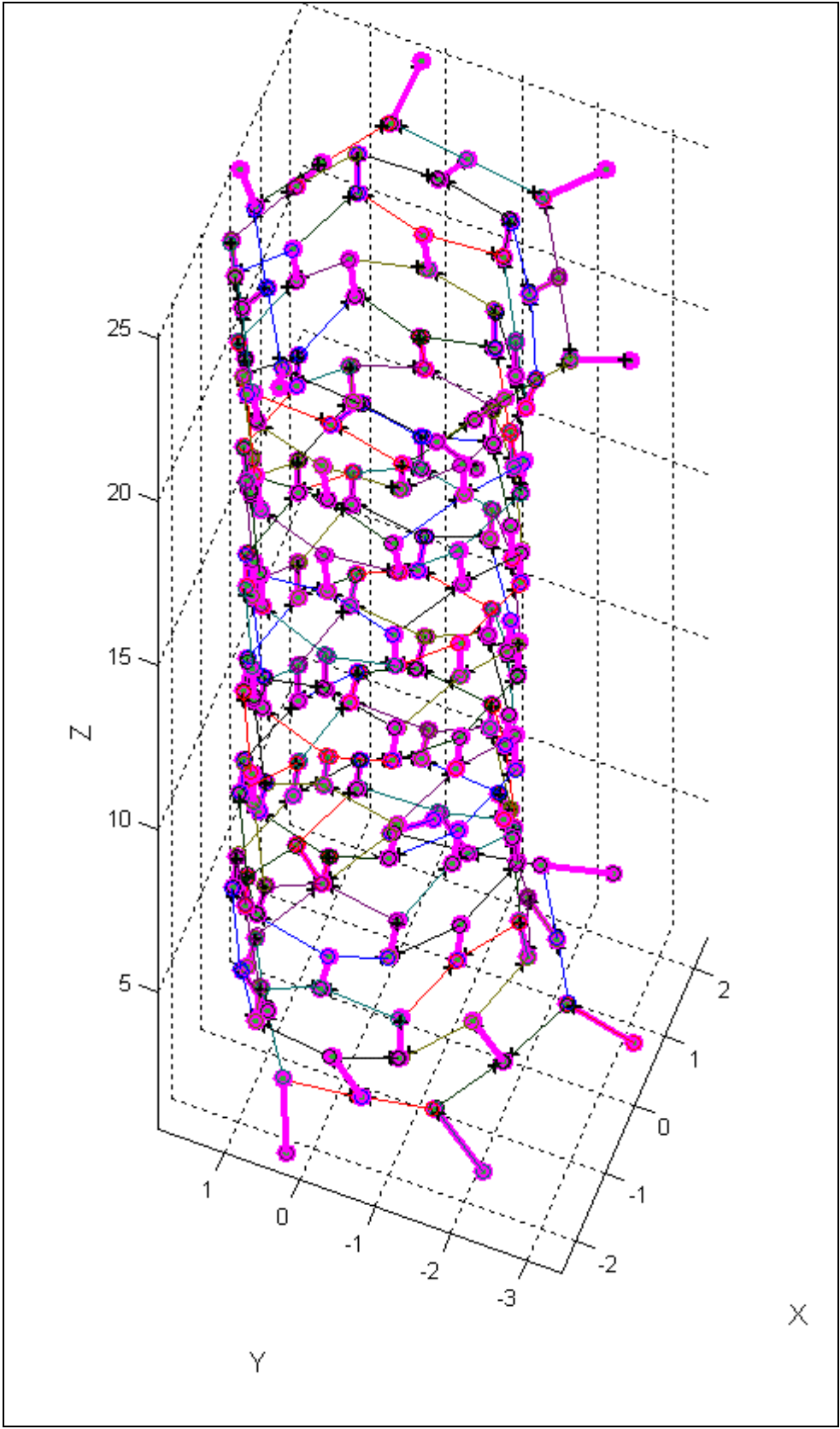

Figure 4. The image of the results of numerical modeling of the behavior of tubular skeletal structure of Fig.2, under conditions listed in Sec. 3, at dimensionless time moment t = 1.0.

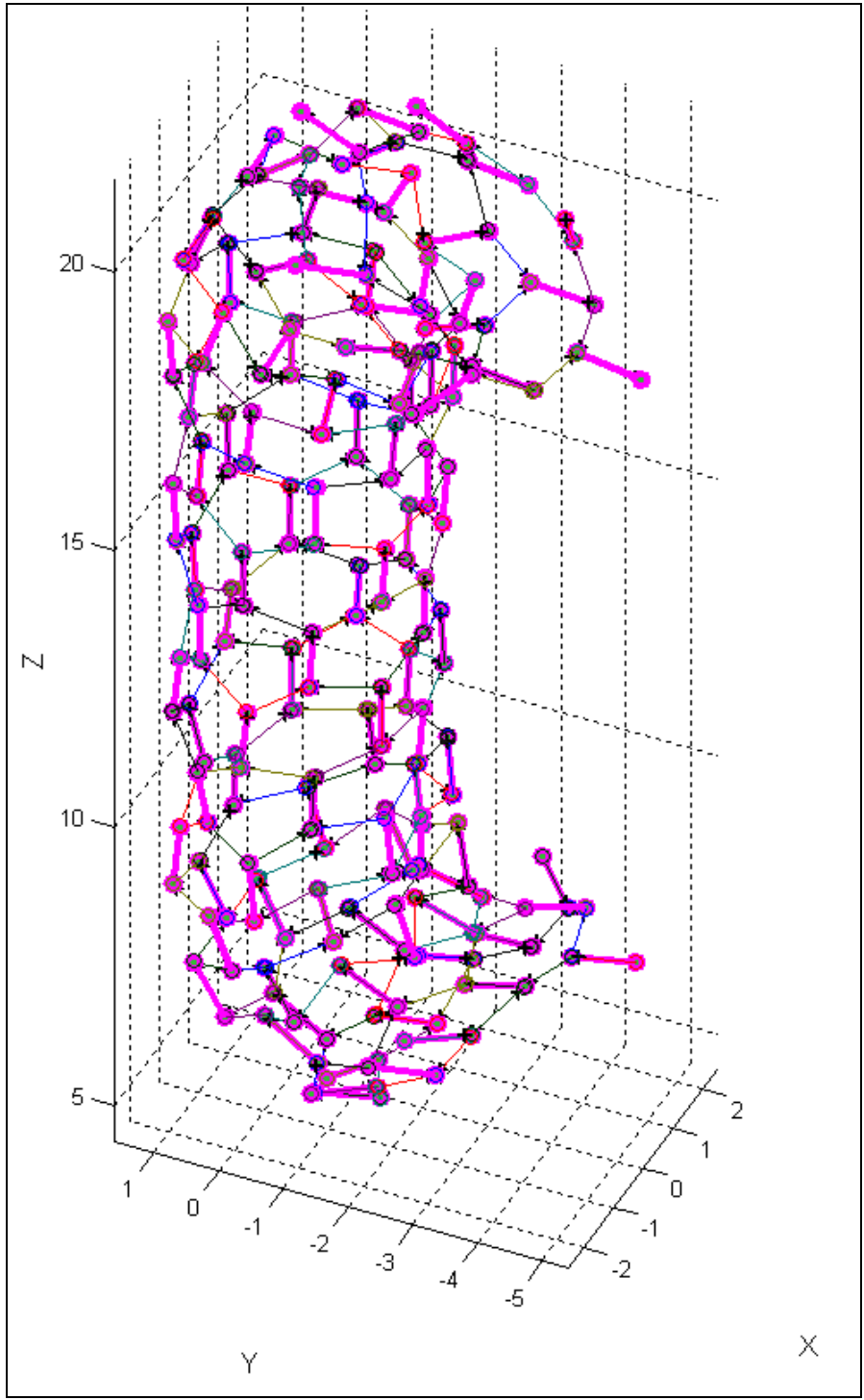

Figure 5. The picture similar to Fig.4, for time t = 2.0.

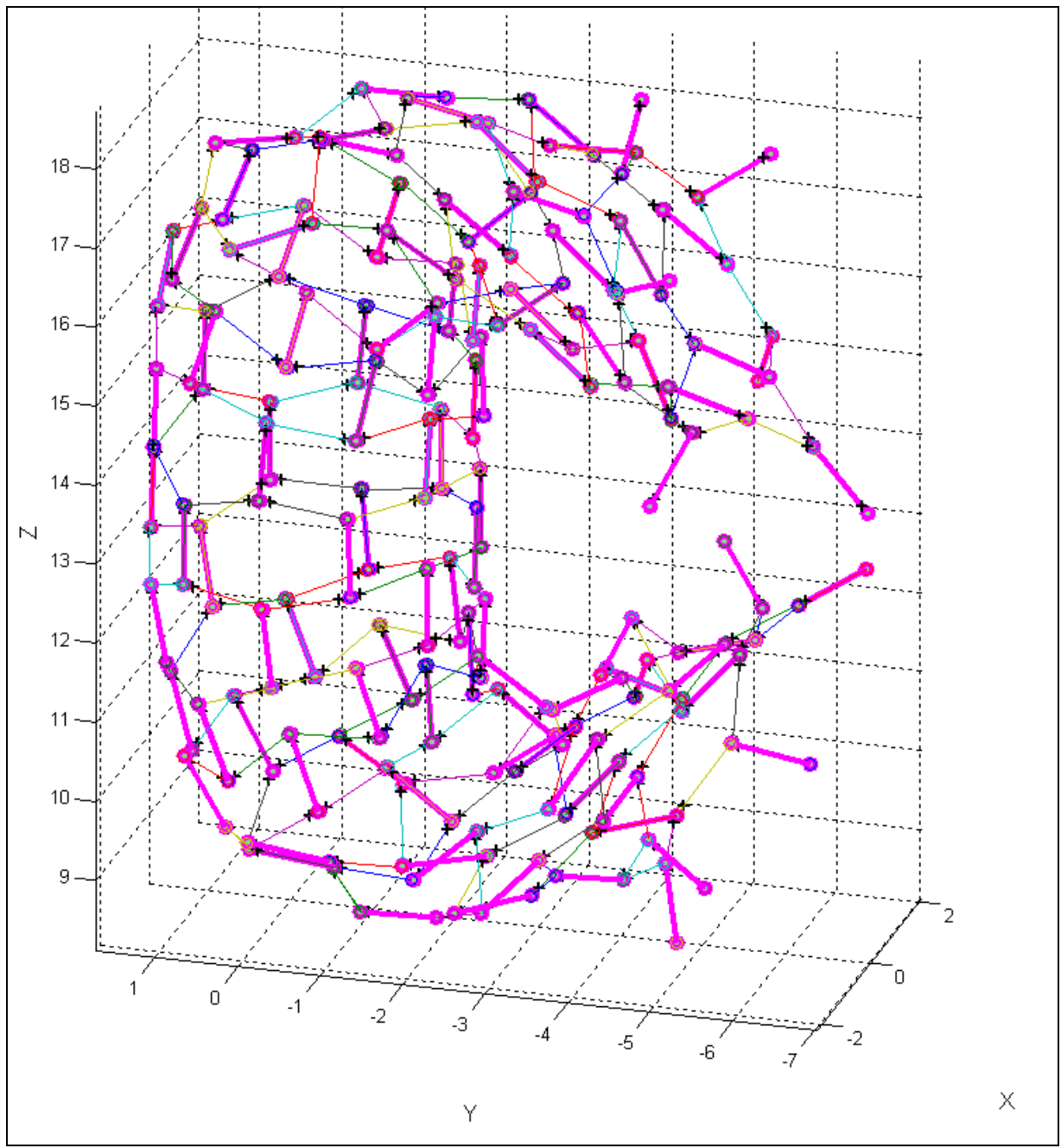

Figure 6. The picture similar to Fig.4, for time t = 2.6.

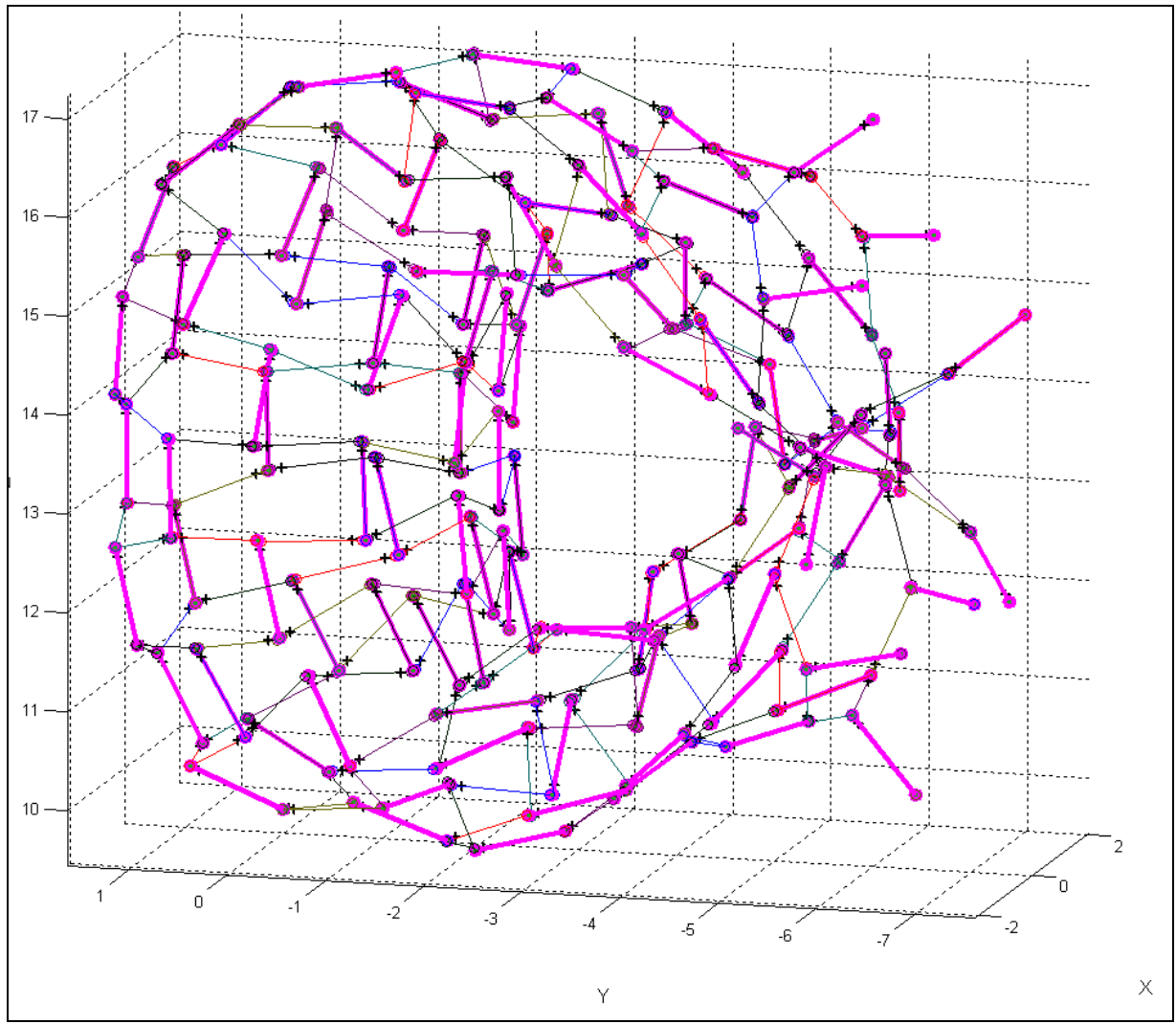

Figure 7. The picture similar to Fig.4, for time t= 2.8.